\documentclass{IEEEcsmag}

\usepackage[colorlinks,urlcolor=blue,linkcolor=blue,citecolor=blue]{hyperref}

\usepackage{xurl}

\usepackage{amssymb} 
\usepackage{upmath}
\usepackage{tikz}
\usetikzlibrary{quantikz}
\jvol{XX}
\jnum{XX}
\paper{X}
\jmonth{XX/XX}
\jname{IEEE Internet Computing}
\pubyear{2022}

\begin{document}

%\sptitle{Department: Head}

%\editor{Editor: Name, xxxx@email}

\title{From Distributed Quantum Computing to Quantum Internet Computing: an Overview}

\author{Seng W. Loke}
\affil{School of Information Technology, Deakin University}

%\markboth{Department Head}{Paper title}

\begin{abstract}
The possibility  of quantum computing has been proposed decades ago, at least as far back as the 1980s, and distributed quantum computing has been studied around two decades ago. Recent times have seen experimental successes and advances in quantum computer hardware and in quantum networking, leading towards the quantum Internet. We provide in this paper an overview of concepts and ideas in distributed quantum computing since over two decades ago as well as look at recent efforts in the area, and consider how, with the development of the quantum Internet, distributed quantum computing is evolving into quantum Internet computing.

\end{abstract}

\maketitle

\chapterinitial{It has been almost twenty years} since the world's first {\em quantum network} became operational, demonstrating quantum key distribution, i.e., the DARPA Quantum Network.\footnote{See \url{https://apps.dtic.mil/dtic/tr/fulltext/u2/a471450.pdf} and \url{https://arxiv.org/pdf/quant-ph/0412029.pdf} [last accessed: 2/8/2022]}   
At the same time, also since around twenty years ago, there have been studies on the communication complexity of quantum distributed system protocols, and how they might compare with classical distributed system protocols, e.g., ~\cite{10.1007/978-3-540-45138-9_1}, as well as work on distributed or non-local quantum gates~\cite{eisert,anocha}, both areas of study have been called {\em distributed quantum computing}. A key resource for many of these quantum protocols and non-local quantum gates is {\em quantum entanglement}, where particular quantum entangled  states are shared by qubits (i.e., two state or or two level quantum-mechanical systems, constituting the basic unit of quantum information) residing on different nodes in a distributed system.

Since then, there have been significant developments in quantum networking, working towards the {\em quantum Internet}, with scale analogous to the (classical) Internet today, with a number of significant demonstrations of quantum networks using fiber optic networks and using free-space (including satellite-based) techniques. Gradually, experimental demonstrations of entanglement over longer and longer distances are being achieved, and an IETF Quantum Network architecture standard is being developed.\footnote{See draft at \url{https://www.ietf.org/id/draft-irtf-qirg-principles-10.html} [last accessed: 3/8/2022]}
Recent work has continued to conceptualise and develop architectures and applications of the quantum Internet~\cite{doi:10.1126/science.aam9288,rohde_2021,cuomo20}. 

A key goal of the quantum Internet is to achieve robust and long distance {\em quantum entanglement} between nodes geographically far apart, e.g., entanglement between qubits on ``any'' two nodes analogous to how two nodes can communicate with each other on the Internet today,  which could then enable  distributed quantum computations and quantum communications among nodes over vast Internet-scale distances.

The aim of this paper is to provide an overview of  {\em quantum Internet Computing}, referring to distributed quantum computing over the emerging quantum Internet. We  look at several key historical developments  as well as review more recent developments in the area of  distributed quantum computing and the quantum Internet.

In particular, in the rest of this paper, we first provide background concepts for this paper, and then provide a conceptual layered overview of quantum Internet computing. Then, we consider the following three  topics  in relation to quantum Internet computing:

\begin{itemize}
\item	distributed quantum computing, which can be divided into two main areas of study (i) quantum protocols involving distributed nodes, from a mainly theoretical perspective involving communication complexity studies and comparisons with classical distributed computing protocols, and (ii) 	distributed quantum computing using non-local or distributed quantum gates, including how a quantum circuit can be distributed across multiple sites for processing.
\item	quantum cloud computing, where we include work on delegating quantum computations from a client to a server presumed to be a universal quantum computer, blind quantum computing where we consider the privacy of client inputs when delegating quantum computations, and verifying delegated quantum computations where we consider how the client can verify if the server has not misbehaved and has perform the delegated computations as wanted by the client, and
\item  computing concerning enabling	quantum Internet including  key ideas such as quantum entanglement distillation, entanglement swapping,  quantum repeaters,  and quantum Internet standards.
\end{itemize}

Finally, we conclude, highlighting the notion of {\em quantum Internet computing}.
% cite from recent special issue: https://ieeexplore.ieee.org/document/9705639 

\section{PRELIMINARIES - A BRIEF BACKGROUND  ON ENTANGLEMENT} 

Central to quantum computing are (i) the ability to manipulate {\em qubits}, instead of bits, i.e. move from certain qubit states to some other qubit states, and (ii) the notion of the {\em qubit}, the basic unit of quantum information, which physically is a two-state (or two-level) quantum-mechanical system such as a photon with vertical and horizontal polarization or electron spin with spin-up and spin-down. The qubit state can be a linear combination of two basis states  $\ket{0}$ and $\ket{1}$, written as $\alpha\ket{0} + \beta \ket{1}$, with probability amplitudes $\alpha$ and $\beta$ where  $|\alpha|^2 + |\beta|^2 = 1$, which when measured becomes the state $\ket{0}$  with probability $|\alpha|^2$ and $\ket{1}$  with probability $|\beta|^2$.

Two qubits $A$ and $B$ can share an entangled state (so-called {\em e-bit}) such as $(\ket{0}_A\ket{0}_B + \ket{1}_A\ket{1}_B)/\sqrt{2}$, or simplified to  $(\ket{00} + \ket{11})/\sqrt{2}$ so that when measured, we have both qubits in the state $\ket{0}$ or both in $\ket{1}$, i.e., there is a correlation that can be used in quantum protocols and algorithms. 

A qubit (its state) can be teleported or transferred from one node to another via transmission of a photon or via entanglement and sending two classical bits. Also, a qubit can be measured resulting in one of the basis states with a probability as determined by the probability amplitude, or undergo a unary operation to transform into possibly a different state. A set of chosen unary operations can form a universal set of quantum gates (e.g., see examples in~\cite{broadbent2015,broadbent2018}), able to represent or approximate all possible unary operations. Quantum gates can be applied to a collection of qubits to perform required computations forming a {\em quantum circuit}, e.g.,  in Figure~\ref{distCNOTeisert}. We discuss later how qubits on one node can affect computations on qubits on another node - using what has been called {\em non-local quantum gates}.

\section{A LAYERED CONCEPTUALIZATION OF  QUANTUM INTERNET COMPUTING}

\begin{center}
\begin{figure}[h]
\includegraphics[width=7cm]{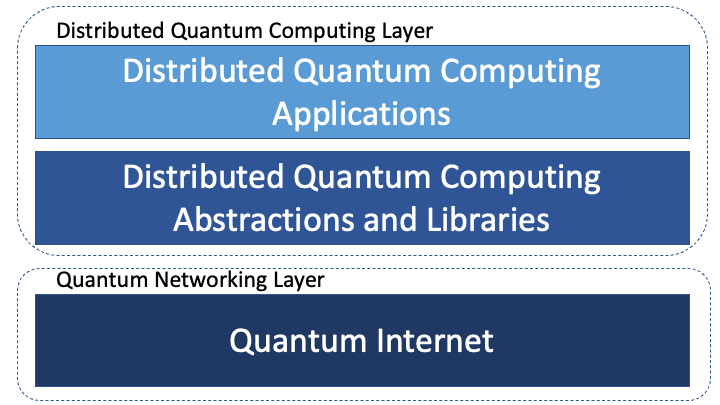}
\caption{Distributed quantum computing over the quantum Internet. Roughly, we have\\
{\em quantum Internet computing} = \\ {\em distributed quantum computing} + {\em quantum Internet}}
\label{arch}
\end{figure}
 \end{center}

As Cuomo {\em et al.} noted in~\cite{cuomo20},  the quantum Internet then becomes the ``underlying infrastructure of the Distributed Quantum Computing ecosystem.'' 
 Figure~\ref{arch} illustrates a three layered conceptualization of the idea of distributed quantum computing over the quantum Internet, where (i) the Distributed Quantum Computing Applications (DQCA) layer comprises the end-user applications using abstractions from the Distributed Quantum Computing Abstractions and Libraries  (DQCA\&L) layer, (ii) the DQCA\&L layer provides a library of code and abstractions to enable end-user applications to be expressed at a high level of abstraction and may perform optimizations to run over the Quantum Internet (QI) layer (e.g., including automatic rewriting of a quantum circuit at the application layer into a collection of (distributed) quantum circuits to run over multiple nodes, and optimizations), and (iii) the QI layer comprises services used for the code generated at the DQCA\&L layer to execute, including preparation of entanglement and transmission of qubits.

\section{DISTRIBUTED QUANTUM COMPUTING}

\subsection{Quantum Advantage for Distributed Computing}
Buhrman {\em et al.}~\cite{buhrman2009} discussed the notion of entanglement and how it could affect the communication complexity of a range of distributed algorithms, showing that some distributed computations can be achieved with fewer bits of communication than is possible classically, and the notion of entanglement enables some distributed computations not possible classically.

As noted by Cleve and Buhrman~\cite{cleve97}, two parties sharing an entangled pair of qubits  cannot use this entanglement to communicate, in the following sense:  suppose parties $A$ and $B$ share an ebit, i.e., they share an entangled pair of qubits (one each), say in the state  $(\ket{00}+\ket{11})/\sqrt{2}$; then when $A$ measures its qubit, and obtains $x=0$ or $x=1$, then $B$ will also obtain the same value, but there is no way $A$ can determine before measurement what $x$ will be, and so, $A$ cannot choose what  $B$ will obtain, i.e. $A$ cannot communicate any bit of information to $B$.  However, a shared ebit can be used to  reduce the communication between $A$ and $B$ when performing some computations, and in fact, with shared entangled qubits among more than two parties, some distributed computation can be achieved with fewer bits of communication, at least asymptotically. For example, in the distributed three party product problem, where one of three parties (say $A$) wants to compute a function of three $n$-bit strings (where each party holds one $n$-bit string), communicating two classical bits among the three parties ($B$ and $C$ each sending one bit to $A$) is sufficient to compute the function, provided an entanglement involving $3n$ qubits (each party holding $n$ qubits), i.e., $n$ triples of qubits, with each triple in an entangled state shared by the three parties in the form 
\[
(\ket{001}+\ket{010} +\ket{100}-\ket{111})/2
\]
But in the classical case without sharing any entangled qubits, it will require the communication of at least three classical bits among the parties for $A$ to compute that function. Hence, it is as though the shared entangled qubits took away the need to share that one classical bit. 

In another problem, the distributed Deutsch-Jozsa problem, from ~\cite{buhrman2009}, two parties, each having an $n$-bit string wants to determine either the two strings are equal or that they differ in exactly $n/2$ positions (and suppose that one or the other must be true),  there is   a quantum solution involving communicating  $log~n$ number of qubits (and recall that for communicating each qubit, two classical bits is required, provided there is a shared ebit, which implies $O(log~n)$ number of classical bits of communication) whereas the best classical solution needs at least $0.007n$ bits to be exchanged, i.e., $O(n)$ number of classical bits of communication; that is, while the  constants would be large, the quantum solution has a much lower communication complexity, for large $n$. 

Another problem, the distributed intersection problem,  where two parties want to compute a slot where they are both available given $n$ slots each, a quantum solution can do so with $O(\sqrt{n})$ qubits of communication, whereas any classical solution would require $O(n)$ classical bits of communication.

The above examples, though rather theoretical, demonstrates how entanglement can reduce communication complexity for certain distributed algorithms, though not for all algorithms. 

Entanglement can also enable distributed computations, not possible classically without entanglement. For example, from \cite{buhrman2009}, suppose we have two nodes $A$ and $B$, each with one input bit, $i_A, i_B \in \{0,1\}$ respectively, and the problem is for $A$ and $B$  to output bits $o_A,o_B \in \{0,1\}$, respectively, such that  the following condition is satisfied:
\[
o_A \oplus o_B = i_A \wedge i_B
\]
with the requirement that they do not communicate their  inputs to each other (i.e., $A$ does not know $B$'s inputs and conversely). There could be multiple rounds of this ``game'' where in each round the inputs are generated uniformly and randomly. 
In each round, $A$ and $B$ {\em win} the game if their outputs satisfy the above condition  and {\em lose}, if not. Classically, the best possible success rate,  after $n$ rounds, is $0.75N$ wins, whereas if $A$ and $B$ share entangled states in each round, they can each perform operations involving the entangled states to get a success rate higher than $0.75N$, as high as $\approx 0.8536N$. In fact, achieving this high number of successes can be a form of test to see if $A$ and $B$ can actually share quantum entanglement and can perform quantum operations - entirely classical $A$ and $B$ will not be able to do so. Futher interesting distributed computations made possible via quantum mechanics are discussed in ~\cite{buhrman2009,10.1145/1412700.1412717}.

\subsection{Quantum Protocols}

At the same time, there has been a great deal of studies on ``quantum versions''\footnote{Fundamentally, the quantum algorithms are quite different from the classical analogues due to the use of entanglement.} of classical distributed computing protocols, including quantum secret sharing, quantum oblivious transfer, quantum key distribution, quantum coin flipping, quantum leader election, quantum anonymous broadcasting, quantum voting, and quantum Byzantine Generals, including pointing out the advantages of quantum information in such protocols.\footnote{See a list at \url{https://wiki.veriqloud.fr/index.php?title=Protocol_Library} [last accessed: 9/8/2022]}
Some of these protocols have been experimentally demonstrated, especially quantum key distribution with products in the market.\footnote{For example, see \url{https://www.idquantique.com/} [last accessed: 9/8/2022]} 

The future will see further development of such quantum protocols. Also,
it is not inconceivable to consider many of these protocols, having been implemented and optimized, become then part of the libraries in the DQCA\&L layer to be used by end-user applications. 

\subsection{Non-Local Quantum Gates}
In parallel to theoretical studies and demonstrations of quantum protocols is work on distributing quantum computations over multiple nodes in the style of typical high performance distributed computing, typified by
the early work on non-local gates~\cite{eisert,anocha}, e.g., a distributed-$CNOT$ gate, where the control qubit is on one node and the target qubit is on a different node. A distributed version of Shor's popular quantum factorization algorithm using non-local gates had been developed as early as 2004~\cite{anocha04}.

Such non-local gates use entanglement shared between nodes as a key resource for the nodes to ``coordinate'' their computations.  To illustrate this idea, Figure~\ref{distCNOTeisert} shows the circuit for a  distributed CNOT gate where $A$ holds the control qubit and $B$ the target qubit (i.e., it is a distributed version of $CNOT(\ket{c}_{A_1} \otimes \ket{t}_{B_2}$) and an entangled state (indicated by the wavy line) between $A$ and $B$ is a resource for performing the distributed CNOT operation. Such a circuit can be extended or modified to have multiple control qubits (on different nodes) or multiple target qubits (on different nodes), or to perform any unitary operation $U$ instead of an $X$ gate operation to have distributed-$U$ gates. It can be shown that this circuit works due to the use of what is called a {\em cat-like} state, of the form $\alpha \ket{00} + \beta \ket{11}$. More specifically, consider the qubits $A_1 A_2 B_1 B_2$ written from left to right, then at slice $1$, the state is:
\[
(\alpha\ket{\underline{0}}_{A_1} \ket{0}_{A_2} \ket{\underline{0}}_{B_1}  + \beta \ket{\underline{1}}_{A_1} \ket{0}_{A_2} \ket{\underline{1}}_{B_1})    \otimes \ket{t}_{B_2} 
\]
And noting the underlined qubits ($A_1$ and $B_1$), they form the cat-like state, i.e. dropping $A_2$, we have 
$(\alpha\ket{\underline{0}}_{A_1}\ket{\underline{0}}_{B_1}  + \beta\ket{\underline{1}}_{A_1} \ket{\underline{1}}_{B_1})    \otimes \ket{t}_{B_2}$
which means that controlling using qubit $B_1$ is effectively controlling using qubit $A_1$ (since both $0$ or both $1$), so that the $CNOT$ gate between slices $1$ and $2$, though between $B_1$ and $B_2$ is effectively between $A_1$ ($\ket{c}$) and $B_2$ ($\ket{t}$), which is what was intended.

\begin{figure*}
\begin{center}
      \begin{quantikz}
   \lstick[2]{A}   A_1=\ket{c}   \qw   & \qw   & \ctrl{1}  & \qw   & \qw & \qw & \qw &          \qw &  \qw   &   \gate{Z} & \qw \ket{c} \\
    A_2=\ket{0}     \qw & \ctrl{}{}  & \targ{}   & \qw & \meter{} &  \cwbend{1} &          &            &  &         \\
    \lstick[2]{B}   B_1= \ket{0}    \qw & \ctrl{}{}  & \qw       & \qw & \qw      & \gate{X}\slice{1}    & \ctrl{1}\slice{2}  & \gate{H} & \meter{} &  \cwbend{-2}       \\
      B_2= \ket{t}   \qw & \qw        & \qw       & \qw & \qw      & \qw         & \targ{}       & \qw        & \qw &  \qw &\qw \ket{t'}
        %\arrow[from=2-3,to=1-3,red,yshift=0.3cm,xshift=0.3cm,thick]{}
      %  \arrow[from=2-2,to=1-2,squiggly,dash,line width=0.1mm]{}
         \arrow[from=3-2,to=2-2,squiggly,dash,line width=0.1mm]{}
    \end{quantikz}
\end{center}
\caption{Distributed CNOT by Eisert et al.~\cite{eisert}}
\label{distCNOTeisert}
\end{figure*}

\subsection{Abstractions and Tools}
More recent work on practical distributed quantum computing have looked at:
\begin{itemize}
\item  {\em automatic quantum circuit partitioning and distribution}: quantum algorithms involving a number of qubits (e.g., from tens to hundreds) would be expressed in very large quantum circuits, and if this is to be partitioned and distributed for processing over multiple nodes, it would be an enormous manual effort; the work in~\cite{PhysRevA.100.032308} views the problem as hypergraph partitioning with the aim of minimising communications between nodes (and so reducing the entanglement resource for the computation).  This problem of distributing a quantum circuit has also been viewed as a balanced $k$-min-cut problem~\cite{DBLP:journals/corr/abs-2206-06437,DBLP:conf/wdag/SundaramGR21} solved in two stages (i) assigning qubits to nodes, and (ii) performing cat-entanglement operations (also called migrations) between nodes (as illustrated earlier).

\item  {\em control and simulation of distributed quantum computations}: the work in~\cite{9651415,9351762} addresses the issue of a control system for a  distributed system of quantum computers, formalizing the notion of parallel programs  over quantum computers and scheduling computations on multiple quantum computers, and providing a simulation environment for networked quantum computers. Multiple QPUs (Quantum Processing Units) can be involved in a computation with QPUs connnected via entanglement, with computations coordinated via a central controller. 
 
\item {\em distributed quantum computing abstractions (and API):} in~\cite{10.1145/3458817.3476172}, an extension of the Message Passing Interface (MPI) is given, called QMPI, to enable implementations of distributed quantum algorithms. QMPI provides primitives to communicate qubits such as {\tt QMPI\_Send} (and the corresponding {\tt QMPI\_Recv}) which teleports a qubit from one node to another, primitives to form entanglement such as {\tt QMPI\_Prepare\_EPR} which prepares an entangled pair of qubits (in state $(\ket{00}+\ket{11})/\sqrt{2}$) between two nodes, and collective operations such as  {\tt QMPI\_Reduce} (and its inverse {\tt QMPI\_Unreduce}) and {\tt QMPI\_Scatter\_move} and {\tt QMPI\_Gather\_move}.
\end{itemize}
Extrapolating from the above developments, in the future, the DQCA\&L layer may have libraries and tools that can allow a program to be expressed using high-level abstractions,\footnote{Although not specifically for distributed quantum computing, Classiq (\url{https://www.classiq.io/}) provides examples of how a quantum circuit can be specified at a high level of abstraction and the corresponding circuit generated.} and then map the program automatically to a collection of (sub)programs on nodes, optimize the subprograms and execute the subprograms, and finally, assemble results consistent with the original program (with execution managed by a controller). Note that control of execution could involve synchronization across nodes, waiting for entanglement to form (using the QI layer),  repeatedly   executing programs, e.g., for the purposes of collecting statistics, or re-running subprograms due to failed runs.

\subsection{Summary}
 
Distributed quantum computing has been developing in the areas above, and we will continue to see new applications and tools for running complex quantum algorithms over distributed notes across the Internet, connected via the QI layer.

\section{QUANTUM CLOUD COMPUTING}
Today, cloud computing has become a standard, flexible and (often) economical way for accessing computing and storage services as well as software services. And a convenient way to  deploy functionality on a controlled access metered basis. 
At the same time, recent quantum computers have been made available by giant tech companies (e.g., Amazon,\footnote{See \url{https://aws.amazon.com/braket/}} IBM,\footnote{See \url{https://quantum-computing.ibm.com/services}}  and Google\footnote{See \url{https://quantumai.google/}}) as well as newer companies (e.g., Rigetti\footnote{See \url{https://docs.rigetti.com/qcs/}} and IonQ\footnote{\url{https://ionq.com/docs}}).\footnote{A list of such companies in quantum cloud computing is here: \url{https://thequantuminsider.com/2022/05/03/13-companies-offering-quantum-cloud-computing-services-in-2022/}.}

This raises questions about trusting a (supposedly universal) quantum computer over the cloud to perform quantum operations as required and protecting the privacy of data when delegating computations (and sending the data) to a remote quantum computer. 
Ways to delegate quantum computations to a remote server by a client with very limited quantum capability (e.g., only able to prepare certain quantum states or data - which are sent to the quantum server to compute with - and perform simple quantum gate operations) have been explored, e.g., see~\cite{broadbent2015}.
Analogous to how homomorphic encryption can be used to protect data, i.e., allowing computing on encrypted data, techniques have been devised to allow quantum computations to be carried out on quantum information that remains encrypted, e.g., using the quantum one-time pad~\cite{broadbent2018,fisher2014}, yielding blind quantum (cloud) computing. 
Also, there have been techniques to not only ensure privacy of delegated data and computations, but also methods to ``test'' the quantum server to ensure or verify that the quantum server behaves, performing the delegated computations as required~\cite{fitzsimons2017,gheorghiu2019}. 

It has been an interesting theoretical question (with practical implications) whether, in general, a classical client can verify the result of a quantum computation (done by a quantum server).  For some problems, it is easy for the client to verify the results returned by the server, e.g., in factorization of a large number,  the client can check if the factors returned are indeed factors of the large number, but in general, it is not always easy to verify the result.  But it can be shown how a  classical client can interact with a quantum server to verify the result of a quantum computation~\cite{doi:10.1137/20M1371828}.

 It remains to see how these methods trade-off against efficiency, and whether we will see practical applications of these methods within the scope of quantum cloud computing. Clients today are entirely classical, and clients normally trust quantum cloud service providers on their offerings. Quantum compute services (with a relatively low number of qubits) is also still the main service since quantum memory is still to be developed, and   other quantum computing algorithms and software services (e.g., quantum Machine Learning, quantum Finance, quantum optimization algorithms, quantum simulation and other applications, e.g., see~\cite{9635678}) have been developed but can be considered still nascent.
 More recently, a service-oriented view of quantum software has been investigated, e.g., see~\cite{moguel22}, where  monitoring and deployment support would be required for using quantum services over the Internet, as well as quality of service considerations (e.g., response times, reliability, availability, costs and so on). Quantum API gateways have been proposed in~\cite{9645184} in order to hide  complexity from clients about deploying quantum software, and helping clients choose the best quantum computer at run-time.

\section{QUANTUM INTERNET}

As mentioned earlier,   entanglement is a key enabler (or resource) for distributed quantum computing and quantum  communications. Achieving robust  teleportation and  entanglement over long distances is one of the key concerns of work on the quantum Internet.  The quantum Internet will need the classical Internet for communications and so will coexist with the classical Internet. The quantum Internet has been discussed at length in~\cite{8910635,9605356,doi:10.1126/science.aam9288,rohde_2021}, including the required protocol stack. We discuss the QI layer only  briefly here.

A challenge is photon loss during transmissions, and errors due to imperfect quantum operations.
Entanglement swapping (ES) has been a well known technique for generating entanglement between nodes given entanglement with intermediate nodes, i.e., to extend the distance over which the entanglement is generated. 
Entanglement purification (EP) is a technique used to increaase the fidelity of entanglement given noisy quantum channels and photon loss. Efforts to provide entanglement between nodes over longer  distances has led to the notion of  quantu repeaters. 
There have been three generations of quantum repeaters (QRs), analogous to but works differently from classical network repeaters ~\cite{jiang16,Yan_2021}; the first generation of QRs use a technique called heralded entanglement generation to deal with photon loss errors and EP to increase the fidelity of entanglement before using ES to generate entanglement over a longer distance. The second generation QRs is similar to the first generation of QRs but use quantum error correction (QEC) techniques instead of  EP   to solve errors due to   imperfect operations, while the third-generation  QRs use QEC techniques to address  both photon loss and operation errors, yielding the highest communication rates. As noted in~\cite{Yan_2021}, for robust  long-distance quantum communications,   longer lived quantum memories, and high-fidelity quantum gates will be needed. 

The late Jonathan Dowling, in~\cite{dowling20}, envisions three generations of the quantum Internet: (i) quantum Internet 1.0: for distributing quantum keys (as already possible with commercial products today)\footnote{A list of companies on quantum encryption (including quantum key distribution) is at: \url{https://thequantuminsider.com/2021/01/11/25-companies-building-the-quantum-cryptography-communications-markets/} [last accessed: 16/8/2022]}, (ii) quantum Internet 2.0: for distributing quantum entanglement  which will make possible wider scale distributed quantum computing, involving star networks with central nodes as   entanglement distribution hubs, and quantum local area networks, where there are clusters of such star networks, and (iii) quantum Internet 3.0: where quantum repeaters are widespread and sufficiently advanced to enable much more scalable quantum networks, with multiple entanglement-sharing routes, which will enable quantum Internet computing at the Internet-scale. Indeed, the QI layer will seek to  provide robust entanglement as a service to the upper layers.
%\footnote{It was also mentioned in~\cite{dowling20} that small low-powered quantum memories, if developed, could  enable quantum connected mobile devices (i.e., a phone has a large number of qubits entangled with qubits on other nodes while charging, and when unplugged, already has a store of such  entangled qubits to use - } 

\section{CONCLUSION - TOWARDS QUANTUM INTERNET COMPUTING}
We envision developments in the coming decades towards a more advanced and robust quantum Internet, and together with the development of  quantum computers, quantum Internet computing can then be realized. Quantum positioning, quantum sensing and quantum federated learning are further examples of  applications. We can conceptualise 
\begin{align*}
& quantum~Internet~computing\\
& = distributed~quantum~computing \\
& \quad + quantum~Internet
\end{align*}
which will continue to develop in all its layers in the coming decades.

\bibliographystyle{plain}
\bibliography{quantum}

\end{document}